\documentclass[9pt, conference]{IEEEtran}
\IEEEoverridecommandlockouts
\usepackage{cite}
\usepackage{amsmath,amssymb,amsfonts}
\usepackage{algorithmic}
\usepackage{graphicx}
\usepackage{textcomp}
\usepackage{xcolor}
\usepackage{caption}
\usepackage{multirow}
\usepackage[utf8]{inputenc}
\usepackage{subcaption}
\def\BibTeX{{\rm B\kern-.05em{\sc i\kern-.025em b}\kern-.08em
    T\kern-.1667em\lower.7ex\hbox{E}\kern-.125emX}}
\begin{document}

\title{Task Scheduling for Efficient Inference of Large Language Models on Single Moderate GPU Systems.}

\author{
    \IEEEauthorblockN{Wenxiang Lin\textsuperscript{1}, Xinglin Pan\textsuperscript{2}, Shaohuai Shi\textsuperscript{1}, Xuan Wang\textsuperscript{1}, Xiaowen Chu\textsuperscript{2}}
    \IEEEauthorblockA{
        \textsuperscript{1}Harbin Institute of Technology, Shenzhen, Shenzhen, China\\
        \textsuperscript{2}The Hong Kong University of Science and Technology (Guangzhou), Guangzhou, China
    }
}

\maketitle
\def \modelname {ScheInfer}
\def \ccname {CC}

\begin{abstract}
Large language models~(LLMs) are known for their high demand on computing resources and memory due to their substantial model size, which leads to inefficient inference on moderate GPU systems. Techniques like quantization or pruning can shrink model sizes but often impair accuracy, making them unsuitable for practical applications. In this work, we introduce \modelname{}, a high-performance inference engine designed to speed up LLM inference without compromising model accuracy. \modelname{} incorporates three innovative methods to increase inference efficiency: 1) model partitioning to allow asynchronous processing of tasks across CPU computation, GPU computation, and CPU-GPU communication, 2) an adaptive partition algorithm to optimize the use of CPU, GPU, and PCIe communication capabilities, and 3) a token assignment strategy to handle diverse prompt and generation tasks during LLM inference. Comprehensive experiments were conducted with various LLMs such as Mixtral, LLaMA-2, Qwen, and PhiMoE across three test environments featuring different CPUs and GPUs. The experimental findings demonstrate that \modelname{} achieves speeds between $1.11\times$ to $1.80\times$ faster in decoding and $1.69\times$ to $6.33\times$ faster in pre-filling, leading to an overall speedup ranging from $1.25\times$ to $2.04\times$ compared to state-of-the-art solutions, llama.cpp and Fiddler.
\end{abstract}


\section{Introduction}
Generative large language models (LLMs) are renowned for their exceptional abilities in many AI applications\cite{DBLP:journals/corr/abs-2407-10671,DBLP:journals/corr/abs-2302-13971,DBLP:conf/nips/BrownMRSKDNSSAA20,Index}. These models are very compute- and memory-hungry due to their large model sizes, so they are mainly deployed in data centers equipped with high-end GPUs (e.g., Nvidia Tesla H100) to provide low-latency and high-throughput services~\cite{aminabadi2022deepspeed}. Recently, it is a burgeoning trend towards running LLMs on more accessible local platforms, such as edge devices~\cite{ye2024asteroid} and personal computers (PCs) with moderate GPUs (e.g., Nvidia RTX 3090)~\cite{josefalbers_2024_13352415,mlc-llm}. This shift is driven by the need for improved data privacy, model customization~\cite{DBLP:conf/naacl/LyuJZXWZCLTL24}, and lower inference expenses~\cite{DBLP:journals/corr/abs-2302-13971}. Deploying LLMs on moderate GPUs poses a challenge because it requires making the model compatible with these moderate GPU systems, and additionally, there is a need to optimize its inference latency to ensure it can handle real-time query processing efficiently.

Current strategies for addressing memory challenges involve model compression and offloading. Compression methods such as quantization~\cite{geng2024quq,parklut2024lutgemm}, distillation~\cite{yuan2023distilling}, and pruning~\cite{ma2023llm} aim to reduce the model size such that the compressed model can be fully loaded to the GPU memory. Yet, even significantly compressed models may still exceed the memory capacity of moderate GPUs, especially on sparse Mixtures of Experts~(MoE) models~\cite{lepikhingshard,jiang2024mixtral}. For example, loading a Mixtral-8x22B MoE model~\cite{jiang2024mixtral} with 4-bit precision requires about 110GB memory for its parameters, surpassing the memory capacity of many moderate GPUs like Nvidia RTX 2080/3090/4090 that have no more than 24GB memory. Model offloading, on the other hand, divides the model across GPU and CPU at the Transformer layer level~\cite{aminabadi2022deepspeed,llamacpp,kamahori2024fiddler}. Leading systems like llama.cpp~\cite{llamacpp} allocate layers between CPU and GPU memory, easing the demand on GPU resources. For MoE models, Fiddler\cite{kamahori2024fiddler} shifts experts to CPU memory to decrease GPU memory needs. Nevertheless, these approaches use either CPU resources (large memory) or GPU resources (high-performance computing) to optimize the inference speed, which is suboptimal.

To this end, this paper presents \modelname{}, an efficient LLM inference system (\S\ref{subsec:system-overview}) designed for local deployment on computer systems with only a single moderate GPU. The main design concept of \modelname{} is to fully utilize the available computing, memory and communication resources of the system. To achieve this goal, \modelname{} partitions the weight tensors (i.e., multi-layer perceptions (MLPs) of dense Transformers or experts in sparse MoE Transformers), which occupy most parameters of the model, into three components, 1) \ccname{}: parameters stored and executed on the CPU, 2) CG: parameters stored on the CPU and executed on the GPU, and 3) GG: parameters stored and executed on the GPU. By doing this, \ccname{}, CG, and GG tasks are possible to be executed simultaneously to fully utilize the available resources of the computer system (\S\ref{subsec:slice} and \S\ref{subsec:taskscheduler}). 


However, the design of \modelname{} still faces notable challenges in achieving optimal performance. \textit{First, how to determine the sizes of \ccname{}, CG, and GG is non-trivial to achieve the minimal inference time due to variations in model size and computer systems.} 
To address this, \modelname{} builds an optimization model~(\S\ref{subsec:profiler-sovler}) to determine the optimal slicing rates which indicate how many parameters should be placed on \ccname{}, CG, and GG. 
\textit{Second, due to the different compute characteristics of the prompt phase and the generation phase during inference, the slicing rates may differ between the two phases. An optimal slicing rate for the generation phase can result in suboptimal performance during the prompt phase.
} We find that the delay arises from matrix-multiplication (GEMM) of the prompt token by the CC matrix taking significantly longer than other operations in the prompt phase with the optimized slicing rate from the generation phase. Therefore, we split the CPU computation task by dividing its input tokens into two parts: 1) remained to do the computation on the CPU, and 2) transferred together with the weight to the GPU for computation (specially denote the weight transferred as CG$'$). The weights are viewed as CC, CG and GG (for tokens processed on the CPU) and as CG$'$, CG, and GG (for tokens processed on the GPU) in the prompt phase, respectively.
Then, we
propose a token assignment strategy (\S\ref{nzero}) that determines the number of tokens run on GPU with CG$'$ and on CPU with CC. We conduct extensive experiments (\S\ref{sec:evaluation}) with four popular LLMs and three representative testbeds and the experimental results demonstrate that \modelname{} runs $1.25\times$ to $2.04\times$ faster than state-of-the-art inference solutions including llama.cpp and Fiddler.


\section{Background and Motivations}
\begin{figure}[!t]
	\centering
\includegraphics[width=0.6\linewidth]{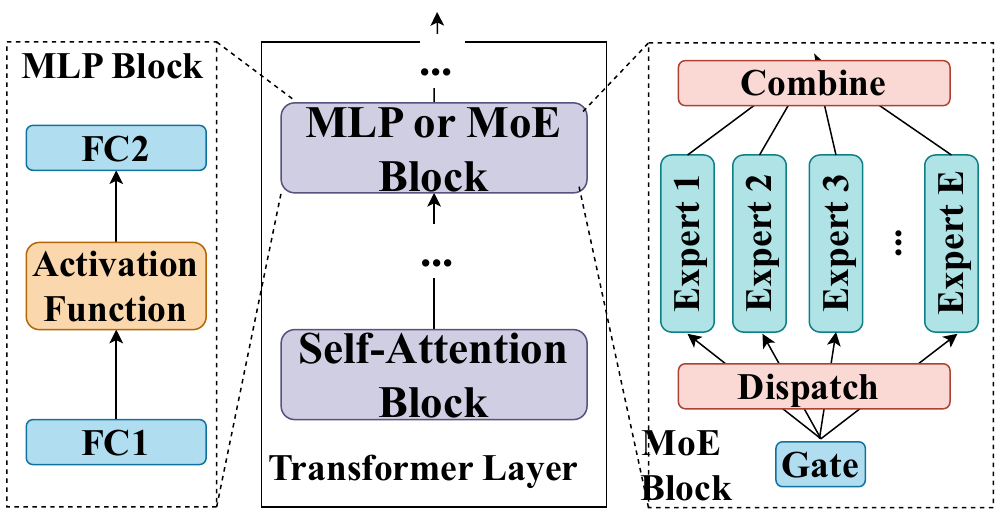}
	\caption{The LLM architecture with dense MLP or sparse MoE blocks.}
 \label{fig:arch}
\end{figure}

\begin{figure}[!t]
	\centering
\includegraphics[width=0.8\linewidth]{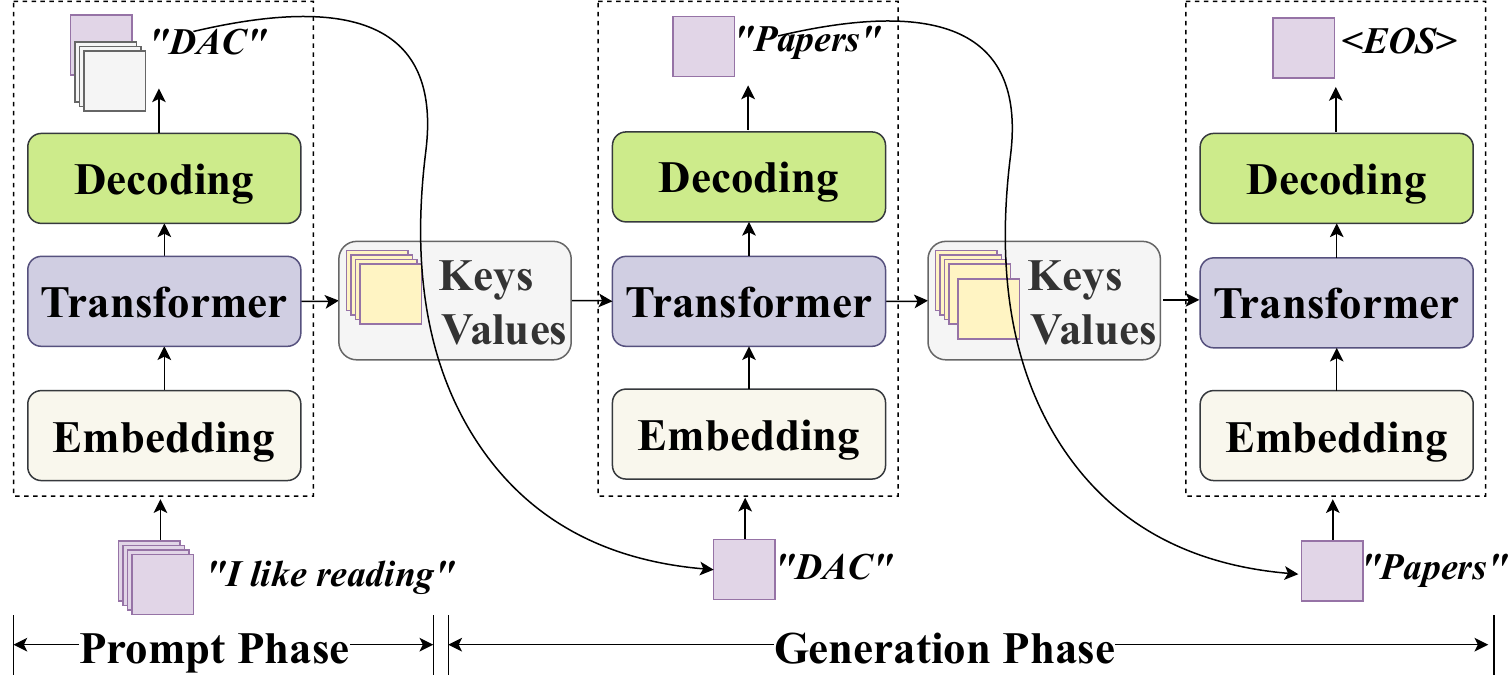}
	\caption{The generative inference procedure of an LLM.}
 \label{fig:LLMinfer}
\end{figure}
\subsection{LLM Architecture and LLM Inference}
Modern LLM architectures mainly consist of multiple Transformer layers, each of which has a self-attention layer and an MLP block (dense LLMs) or an MoE block (sparse MoE LLMs) as shown in Fig.~\ref{fig:arch}. The self-attention layer builds input sequence representations by identifying relationships between tokens. In contrast, the MLP or MoE block uses fully connected layers with activation functions to enhance the input sequence representation. Specifically, the dense MLP block processes the input by two fully connected layers (FC1 and FC2) with an activation function in the middle~\cite{vaswani2017attention}. The MoE block employs a gating function to assign tokens to their respective experts, subsequently processing each token with their experts and aggregating the results~\cite{lepikhingshard}.

A typical generative inference process in large language models involves two stages: 1) the prompt phase and the generation phase. The prompt phase is designed to create a KV cache for each layer. The input of this phase is a prompts, such as a lengthy instruction sentence. 2) The generation phase is iterative, updating KV caches and generating tokens step-by-step. The key difference between these phases is that the generation phase is executed multiple times although the input sequence for each iteration is one.

\subsection{Motivations}
\textbf{Insufficient utilization of limited resources.}
Consumer-grade computers often have limited GPU memory and CPU computational resources. Maximizing these available resources to enhance inference performance presents a significant challenge. Prior studies typically exploit just one type of resource at a time. For instance, Fiddler~\cite{kamahori2024fiddler} optimizes the inference of MoE LLMs by solely utilizing CPU resources for MoE layers and solely exploiting GPU resources for other Non-MoE layers. Llama.cpp~\cite{llamacpp}, instead, partitions the model layers between the CPU and GPU, processing them on their respective processors. Mixtral-Offloading~\cite{eliseev2023fast} involves loading expert weights onto the GPU for computation. \textit{However, none of these approaches simultaneously exploit multiple resources, leading to inefficiencies.} Consequently, our approach aims to support parallel execution of CPU and GPU tasks, along with efficient communication between CPU and GPU, to maximize resource usage. 
Inspired by tensor parallelism~\cite{DBLP:dean2012large}, we propose to partition weight tensors into multiple parts to be executed on CPU and GPU simultaneously.

\textbf{Determining the slicing rate of each part.}
Though we enable the simultaneous execution on both CPU and GPU, how to determine how many parameters (i.e., the optimal slicing rate) should be placed on CPU or GPU is challenging the due to varied model sizes and different computing capability of CPU and GPU. Our experimental results demonstrate that the optimal slicing rate can be up to $2.1\times$ faster on average than a manually configured slicing rate (\S\ref{subsubsec:abla_rate}). Notably, the optimal slicing rate varies significantly due to its heavy reliance on the computational and storage capacities of consumer-grade devices, which can differ markedly. We summarize its differences into two challenges: 1) The computational and memory capacities of CPUs and GPUs vary. Thus, unlike homogeneous GPUs with identical capacities, we cannot employ tensor parallelism in the same way. 2) There is a large variation in computational and memory capacities among different CPUs, GPUs and the interconnect between the CPU and GPU. Thus, determining the best slicing rate needs to be tailored to each specific consumer-grade computer.



\textbf{Different workloads between Prompt and Generation Phases.}
Because the workloads for the prompt and generation phases are significantly different, the best slicing rate differs to achieve the best inference performance. \textit{That means we must frequently merge CC, CG, and GG tensors from both CPU and GPU memory, reorganize them into contiguous and executable weights, and distribute them to appropriate locations in the prompt phase in order to utilize the optimal slicing rates.} 
Normally, the generation phase takes much more time than the prompt phase. Thus, we fix the slicing rates to those optimized in the generation phase, ensuring that the generation phase can, at the very least, use the optimal slicing rates. However, this will result in poor performance in the prompt phase. To alleviate the issue, with the finding that CPU computation takes much longer than other operations, we propose to adjust the number of tokens executed on CPU (i.e. executed with CC tensors). To achieve this, we additionally introduce CG$'$ parameters in the prompt phase which share the same memory addresses with CC but will be executed on GPU. Then, we reuse the optimization model to determine slicing rates to automatically determine the number of tokens that run on CPU with CC and on GPU with CG$'$ to optimize the performance in the prompt phase. 

\begin{figure}[!ht]
	\centering
\includegraphics[width=\linewidth]{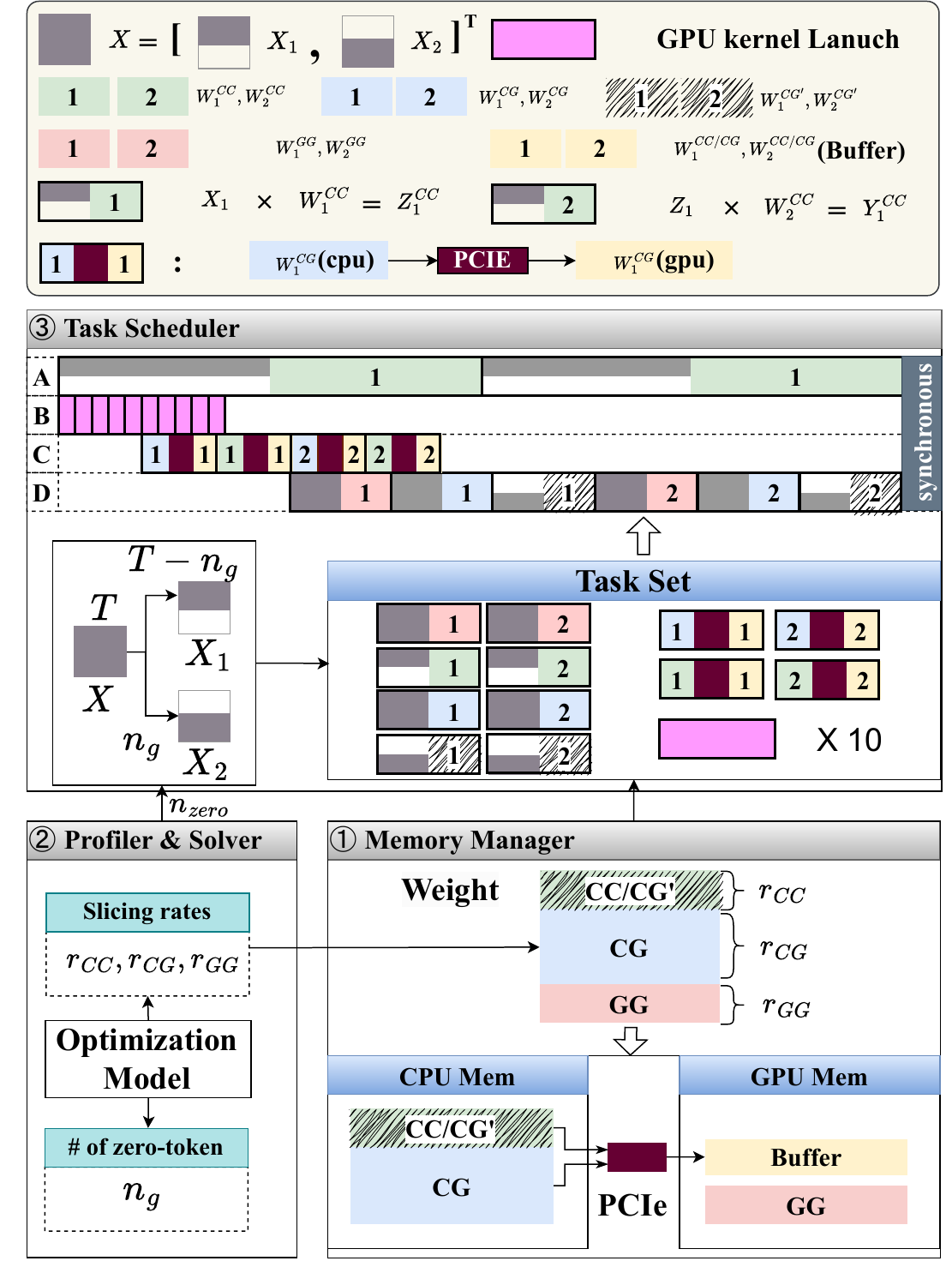}
	\caption{The system overview of \modelname{} with an example of two linear layers in an MLP block.}
 \label{fig:engine}
\end{figure}

\begin{figure}[!ht]
	\centering
\includegraphics[width=\linewidth]{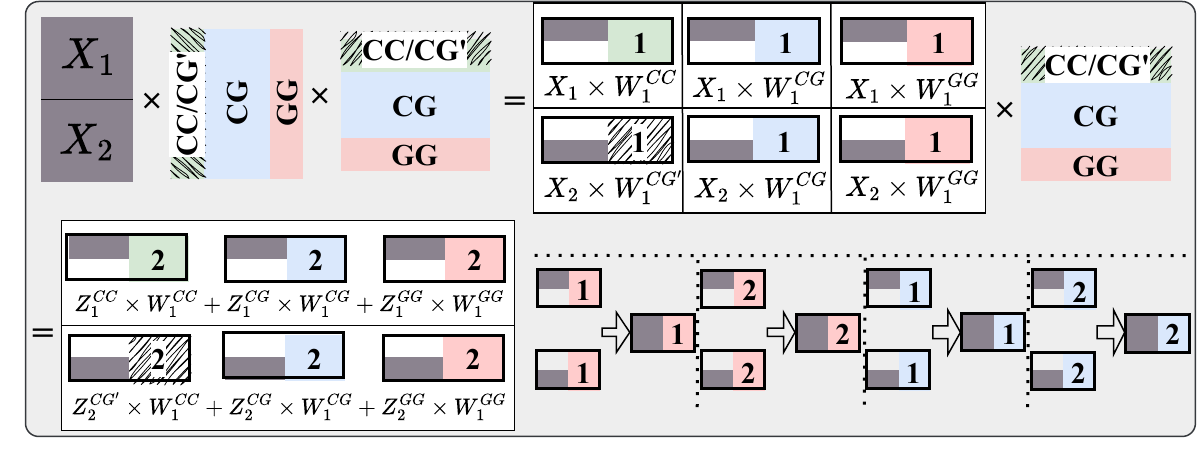}
	\caption{The illustration of slicing input tokens and weights.}
 \label{fig:tp}
\end{figure}

\section{System Design of \modelname{}}
\subsection{System Overview}\label{subsec:system-overview}
To better utilize different resources, we propose our \modelname{} that slices the weight tensors of MLP or MoE layers into three parts to enable the parallel execution of different resources (CPU, GPU, and PCIe interconnect between CPU and GPU). To support the parallel execution, the design of our \modelname{} consists of three components, including Memory Manger, Profiler \& Solver, and Task Scheduler as shown in Fig.~\ref{fig:engine}.
First, Memory Manager manages to slice the weight tensors and supervises the allocation of their storage so as to handle the memory addresses required for transferring weight tensors from the CPU to the GPU. Second, Profiler \& Solver involve solving the optimal slicing rates by profiling running-time information for any given models and hardware. Third, Task Scheduler organizes the execution of the MLP or MoE layers and ensures maximum parallelism by coordinating CPU calculations, GPU computations, GPU kernel launches, and GPU-CPU communications.

\subsection{Memory Manager: Slicing Weight Tensors}\label{subsec:slice}
In Memory Manager, the weight tensors of an MLP or MoE layer are divided into three parts including 1) CC: parameters are stored and executed on the CPU side, 2) CG: parameters are stored on the CPU side while executed on the GPU side, and 3) GG: parameters are stored and executed on the GPU side as shown in Fig.~\ref{fig:engine} (Memory Manager). Additionally, Memory Manager will maintain the CG$'$ parameters for the prompt phase which share the same memory addresses with CC but will be executed on the GPU side. For CC tensors, they are executed on CPU (requiring CPU resources) and their results are then transferred to GPU memory as the inputs of their next layers (requiring the PCIe resource). For CG and CG$'$ tensors, they are stored on CPU memory to save GPU memory, but they need to be transferred to GPU memory for execution (requiring the PCIe resource and the GPU computing resource). Generally, CPU computation is slow, so CG tensors utilize PCIe and GPU resources to decrease the CPU's workload, thereby enhancing efficiency. For GG tensors, they are executed on GPU (requiring the GPU computing resource). With these three tensor types, we can simultaneously leverage CPU, GPU, and PCIe resources.

Fig.~\ref{fig:tp} shows an example of our weight slicing in an MLP layer with two linear layers, where we ignore the activation computation for better presentation. Formally, let $X \rightarrow [X_{1}\ X_{2}]^{\top}$ represent the input tensor stored on GPU of the MLP layer with $T$ tokens. Notably, $X_{1}$ will be additionally transferred into CPU (ignorable cost) to simultaneously calculate with tensors on CPU and GPU. And $X_{2}$ is divided by $n_{g}$ to reduce CPU computation in the prompt phase (detailed in \S\ref{nzero}).
$W_1, W_2$ represent the weight tensors of FC1 and FC2 in the MLP layer, respectively. $W_1$ and $W_2$ are sliced into CC, CG, and GG, that is $W_1 \rightarrow [W_1^{\ccname{}}\ W_1^{CG}\ W_1^{GG}]$ and $W_2\rightarrow [{W_2^{\ccname{}}}\ {W_2^{CG}}\ {W_2^{GG}}]^\top$. 
The computation of the MLP layer can be divided as:
\begin{equation}
\begin{bmatrix} Z_1^{\ccname{}} & Z_1^{CG} & Z_1^{GG} \\ Z_2^{CG'} & Z_2^{CG} & Z_2^{GG} \end{bmatrix} = A(\begin{bmatrix} X_1 \\ X_2 \end{bmatrix} \begin{bmatrix} W_1^{\ccname{}} & W_1^{CG} & W_1^{GG} \end{bmatrix} ),
\end{equation}
\begin{equation}
\begin{bmatrix} Y_1^{\ccname{}}+Y_1^{CG}+Y_1^{GG} \\ Y_2^{CG'}+Y_2^{CG}+Y_2^{GG} \end{bmatrix}  = \begin{bmatrix} Z_1^{\ccname{}} & Z_1^{CG} & Z_1^{GG} \\ Z_2^{CG'} & Z_2^{CG} & Z_2^{GG} \end{bmatrix} \begin{bmatrix}  W_2^{\ccname{}} \\ W_2^{CG} \\ W_2^{GG}\end{bmatrix}, 
\end{equation}
Here, $A(\cdot)$ denotes the activation function applied element-wise. As $X_{2}$ will be executed with CG$'$ instead of CC, we substitute $Z_{2}^{CC}=X_{2} \times W_{1}^{CC}$, $Y_{2}^{CC}=Z_{2}^{CC} \times W_{2}^{CC}$ with $Z_{2}^{CG'}=X_{2} \times W_{1}^{CG'}$, $Y_{2}^{CG'}=X_{2} \times W_{2}^{CG'}$.
The CPU handles calculations for $Z_1^{CC}$ and $Y_1^{CC}$, while the GPU processes the remaining computations (Weight tensors stored on the CPU will be transferred to the GPU for execution.). 

In comparison to the unpartitioned original $Y_1=W_2(A(W_1\times X_1))$ and $Y_2=W_2(A(W_1\times X_2))$, the output remains identical after concatenating $Y_1^{\ccname{}}+Y_1^{CG}+Y_1^{GG}$ and $Y_2^{CG'}+Y_2^{CG}+Y_2^{GG}$ into a final result, akin to Tensor Parallel.

\subsection{Task Scheduler: Scheduling Tasks with Different Streams}\label{subsec:taskscheduler}
Task Scheduler provides the capability for executing different types of tasks in parallel and enhances the performance by arranging the CPU computation tasks, GPU computation tasks, and the communication tasks between the CPU and GPU to maximize their overlaps. Be aware that the GPU computation task includes a GPU kernel launch time, which might be similar in duration to the GPU's computational time when only a few tokens are processed during the generation phase. Fig.~\ref{fig:engine} (Task Scheduler) provides an example of our scheduler managing an MLP block (the MoE block functions similarly, as the key difference between MLP and MoE lies in the number of matrix-multiplication or GEMM operations) during the prompt phase with two linear layers.
We organize CPU computation tasks, GPU computation tasks, and CPU-GPU communication tasks into a task set for efficient scheduling. Fig.~\ref{fig:tp} illustrates the outcomes of tasks from the sets depicted in Fig.~\ref{fig:engine}. Notably, $X_{2}$ allocated $n_{g}$ tokens from $X$ with $T$ tokens is only to manage CPU computation during the prompt phase, ensuring no impact on GPU computation. Therefore, as shown in Fig.~\ref{fig:tp} (right lower), we combine $X_{1} \times W_1^{CG/GG}$ and $X_{2} \times W_1^{CG/GG}$ as $X \times W_1^{CG/GG}$ as well as $Z_{1}^{CG/GG} \times W_2^{CG/GG}$ and $Z_{2}^{CG/GG} \times W_2^{CG/GG}$ as $Z^{CG/GG} \times W_2^{CG/GG}$. CPU computation tasks involve $X_1 \times W_1^{CC}$, $Z_1^{CC} \times W_2^{CC}$. GPU computation tasks involve $X \times W_1^{CG}$, $Z^{CG} \times W_2^{CG}$, $X \times W_1^{GG}$, $Z^{GG} \times W_2^{GG}$, $X_2 \times W_1^{CG'}$, $Z_2^{CG'} \times W_2^{CG'}$.  Communication tasks include transferring $W^{CG}$ and $W^{CG'}$ from the CPU to the GPU. 
During the generation phase, $n_{g}$ is set to zero, and any tasks associated with $X_2$ will be omitted.

With so many tasks, we set up four asynchronous execution streams in the pipelining (independent operations from different streams can be carried out simultaneously), including CPU computation (Stream-A), GPU kernel launch (Stream-B), communication between CPU and GPU (Stream-C) and GPU computation (Stream-D). Additionally, launching a GPU kernel demands minimal computational power, allowing it to proceed concurrently with CPU operations. The \ccname{} part is executed in Stream-A, the GG part requires both Stream-B and Stream-D, and the CG part requires Stream-B, C and D. There are two types of dependence among these streams. GPU computation for CG can be executed only after their weight communication completes. Weight communication and GPU computation can be executed only after their corresponding GPU kernel launches complete. 

\subsection{Profiler \& Solver: Optimizing Slicing Rates}\label{subsec:profiler-sovler}
For a given model and a testbed, Profiler \& Solver first profiles some important parameters to build performance models, and then solves the best rates for \ccname{}, CG, and GG (\S\ref{subsec:profiler-sovler} will introduce a general optimization model to determine slicing rates with an intact input tensor without splitting). In addition, this module also solves the number of tokens to run with CG$'$ ($n_{g}$) to reduce CPU computation in the prompt phases (\S\ref{nzero} will present a modification to the optimization model to address $n_{g}$.). As shown in Fig.~\ref{fig:arch} (Profiler \& Solver), it generates the best slicing rates for Memory Manager and the optimal \# of tokens to run with CG$'$ for Task Scheduler.

\begin{figure}[!t]
	\centering
	\begin{subfigure}[b]{0.45\textwidth}
		\centering
		\includegraphics[width=\linewidth]{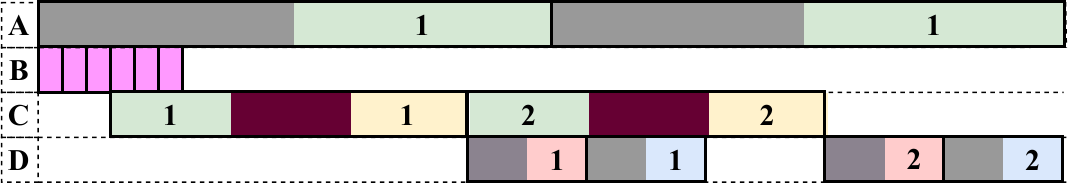}
		\caption{Case1}
		\label{fig:case1}
	\end{subfigure}

	\begin{subfigure}[b]{0.45\textwidth}
		\centering
		\includegraphics[width=\linewidth]{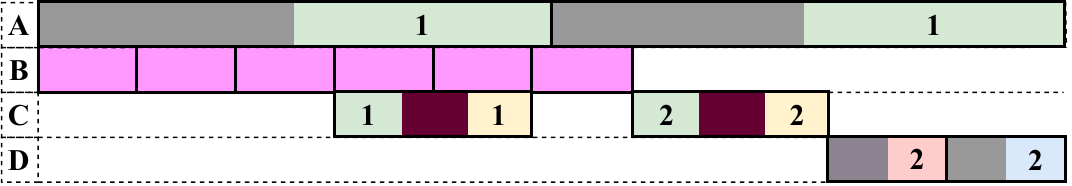}
		\caption{Case2}
		\label{fig:case2}
	\end{subfigure}

 \begin{subfigure}[b]{0.45\textwidth}
		\centering
		\includegraphics[width=\linewidth]{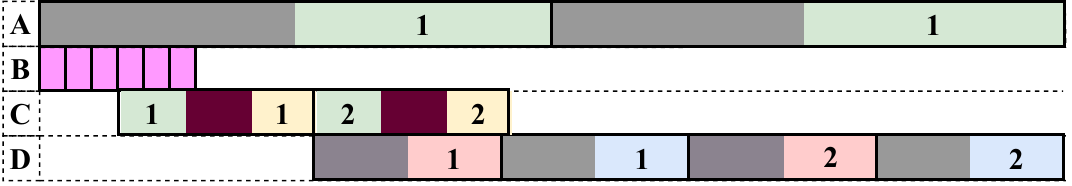}
		\caption{Case3}
		\label{fig:case3}
	\end{subfigure}
	
	\caption{Three schedule cases. The legend is the same as Fig.~\ref{fig:engine}.}
	\label{fig:cases}
\end{figure}

\section{Optimizing Algorithms in Profiler \& Solver}\label{sec:algo-profiler-sovler}

\subsection{Algorithm of Optimizing Slicing Rates}\label{subsec:profiler-sovler}
To achieve the best slicing rates (i.e., $r_{CC}, r_{CG}$, and $r_{GG}$) for partitioning weight tensors, we develop a two-stage optimization algorithm. In the first stage, we solve $r_{CC}$ and  $r_{CG}$ by fixing $r_{GG}$ which is mainly related to GPU memory. In the second stage, we restrict the search space of $r_{GG}$ into integers and calculate the time cost bonus for each integer using the model from the first stage, which solves the best schedule to assign GPU memory with a greedy algorithm. Notably, optimizing slicing rates will be operated only once.

\subsubsection{Performance Models}\label{sec4.1}
Le $t_{G}$, $t_{C}$, $t_{C2G}$, and $t_{Launch}$ denote the time taken for a GPU GEMM operation, a CPU GEMM operation, data transfer from CPU to GPU, and a GPU kernel launch, respectively.
To simply the problem, we model $t_{Launch}$ as a constant and $t_{G}$, $t_{C}$, $t_{C2G}$ as linear models~\cite{shi2023pipemoe}, i.e.,
\begin{equation}\label{pm}
\left\{    \begin{array}{ll}
{t_{G}} & = {\alpha _{G }} + {n_{G}} \cdot {\beta _{G}},\\
{t_{C}} & = {\alpha _{C }} + {n_{C}} \cdot {\beta _{C}},\\
{t_{C2G}} & = {\alpha _{C2G}} + {n_{C2G}} \cdot {\beta _{C2G}},\\
{t_{Launch}} & = \text{constant},
\end{array}\right.
\end{equation}
where $n_{*}$ represents the volume of the communication message or the workload of GEMM (i.e., dimensions of two input matrices), ${\alpha _{*}}$ denotes the startup time and ${\beta _{*}}$ represents the time per byte transmitted or per unit of workload processed. 

\subsubsection{Problem Formulation}
During the model inference, $n_{GEMM}=T\cdot M \cdot H$ where $T$, $M$, and $H$ denote the number of tokens, the model dimension, the hidden dimension, respectively. The shapes of the input tensor and the weight tensor are $[T,M]$ and $[M,H]$, respectively.
Let $n_W$ denote the total bytes of the weight tensor, then we have $n_{G,GG}=r_{GG} \cdot n_{GEMM}$, $n_{C,\ccname{}}=r_{\ccname{}} \cdot n_{GEMM}$ , $n_{G,CG}=r_{CG} \cdot n_{GEMM}$ and $n_{C2G,CG}=r_{CG} \cdot n_{W}$. 
According to Eq.~\ref{pm}, we obtain
\begin{equation}
\left\{ \begin{array}{ll}
    {t_{G}} &= {\alpha _G} \cdot [{\mathop{ sgn}} ({r_{CG}}) + {\mathop{ sgn}} ({r_{GG}})] \\
    &\ + ({r_{CG}} + {r_{GG}}) \cdot {n_{GEMM}} \cdot {\beta _G},\\
{t_{C}} &= {\alpha _C} \cdot {\mathop{ sgn}} ({r_{\ccname{}}}) + {r_{\ccname{}}} \cdot {n_{GEMM}} \cdot {\beta _C},\\
{t_{C2G}} &={\alpha _{C2G}} + {r_{CG}} \cdot {n_W} \cdot {\beta _{C2G}},\\
{t_{L}} &= [2\cdot {\mathop{ sgn}} ({r_{CG}}) + {\mathop{ sgn}} ({r_{GG}})]\cdot t_{Lanuch},
\end{array} \right.
\end{equation}
where $sgn(\cdot)$ denotes the sign function. Notably, a CG operation requires a launch to transfer data from CPU to GPU and a launch to execute GPU computation, so its coefficient is 2.

Let $\tau _G^{i}$, $\tau _C^{i}$, $\tau _{C2G}^{i}$ and $\tau _L^{i}$ denote the completion timestamp of CPU GEMM computation, communication from CPU to GPU and the GPU kernel launch for the $i_{th}$ GEMM computation in an MLP or MoE layer. Thus, the dependency of these tasks can be formally described as 
\begin{equation}\label{timestamp}
\left\{ \begin{array}{ll}
\tau _L^i &= \tau _L^{i - 1} + {t_L},\\
\tau _{C2G}^i &= \max (\tau _L^i,\tau _{C2G}^{i - 1}) + {t_{C2G}},\ 0 \le i \le {n_l},\\
\tau _G^i &= \max (\tau _{C2G}^i,\tau _G^{i - 1}) + {t_G},\\
\tau _C^i &= \tau _C^{i - 1} + {t_C},
\end{array} \right.
\end{equation}
where $n_{l}$ denotes the number of GEMMs in an MLP or MoE layer. 
And the time cost of the MLP or MoE layer is represented as $t_{fin}=max(\tau_{G}^{n_{l}},\tau_{C}^{n_{l}})$.
Thus, our goal is to minimize $t_{fin}$ by changing the slicing rates. In the first stage, we only optimize $r_{CG}, 0\leq r_{CG} \leq (1-r_{GG})$ since $r_{GG}$ is related to the GPU memory bound (discussed in \S\ref{subsub:gpumem}), and $r_{\ccname{}}=1-r_{CG}-r_{GG}$. 
\subsubsection{Optimal Solution}
According to Eq.~\ref{timestamp}, $\tau_{C}^{n_l}=n_{l}\cdot t_{C}$ and we eliminate the max functions by using the following conditions.
\begin{equation}
\begin{array}{c}
        Q1: {t_{L}} < {t_{C2G}},  
\end{array}
\end{equation}
\begin{equation}
    \begin{array}{c}
         Q2: {t_G} < {t_{C2G}},
    \end{array}
\end{equation}
\begin{equation}
    \begin{array}{c}
         Q3:{t_{L}} < t_{G}.
    \end{array}
\end{equation}
Under these conditions, we can categorize all scenarios into three distinct cases.

\textbf{Case1 (Q1 is true and Q2 is true)}: It indicates the communication time between CPU and GPU is larger than the computation time of GPU GEMM and the launch time of GPU kernels. Thus, the communication between CPU and GPU dominates the overall time cost as shown in Fig.~\ref{fig:case1}. So we obtain
\begin{equation}
\tau _L^i \le \tau _{C2G}^{i - 1} \text{ and } \tau _G^{i - 1} \le \tau _{C2G}^i,
\end{equation}
to eliminate the max function in Eq.~\ref{timestamp}, resulting in
\begin{equation}
    \begin{array}{ll}
\tau _G^{{n_l}} &= {t_{L}} + {n_l} \cdot {t_{C2G}} + {t_G}\\
 &= {t_{L}} + {n_l} \cdot ({\alpha _{C2G}}\cdot {\mathop{ sgn}} ({r_{CG}}) + {r_{CG}} \cdot {n_W} \cdot {\beta _{C2G}})\\
 &\ + {\alpha _G} \cdot {\mathop{ sgn}} ({r_{CG}}) + {r_{CG}} \cdot {n_{GEMM}} \cdot {\beta _G}\\
 &\ + {\alpha _G} \cdot {\mathop{ sgn}} ({r_{GG}}) + {r_{GG}} \cdot {n_{GEMM}} \cdot {\beta _G}.
\end{array}
\end{equation}

\textbf{Case2 (Q1 is true and Q2 is false) or (Q1 is false and Q3 is true)}: It indicates that the GPU computation dominates the overall time cost as shown in Fig.~\ref{fig:case2}. So we have
$\tau _{C2G}^i \le \tau _G^{i - 1}$.
Then we can obtain
\begin{equation}
\begin{array}{ll}
\tau _G^{{n_l}} &= {t_L} + {t_{C2G}} + {n_l} \cdot {t_G}\\
 &= {t_L} + {\alpha _{C2G}}\cdot {\mathop{ sgn}} ({r_{CG}}) + {r_{CG}} \cdot {n_W} \cdot {\beta _{C2G}}\\
 &\ + {n_l} \cdot {\alpha _G} \cdot {\mathop{ sgn}} ({r_{CG}}) + {n_l} \cdot {r_{CG}} \cdot {n_{GEMM}} \cdot {\beta _G}\\
 &\ + {n_l} \cdot {\alpha _G} \cdot {\mathop{ sgn}} ({r_{GG}}) + {n_l} \cdot {r_{GG}} \cdot {n_{GEMM}} \cdot {\beta _G}.
\end{array}
\end{equation}

\textbf{Case3 (Q1 is false and Q3 is False)}: It indicates that the launch time of GPU kernels dominates the overall time cost as shown in Fig.~\ref{fig:case3}. So we have
$\tau _{C2G}^{i - 1} \le \tau _L^i,\tau _G^{i - 1} \le \tau _{C2G}^i$.
Then we can obtain
\begin{equation}
    \begin{array}{ll}
\tau _G^{{n_l}} &= {n_l} \cdot {t_L} + {t_{C2G}} + {t_G}\\
 &= {n_l} \cdot {t_L} + {\alpha _{C2G}}\cdot {\mathop{ sgn}} ({r_{CG}}) + {r_{CG}} \cdot {n_W} \cdot {\beta _{C2G}}\\
 &\ + {\alpha _G} \cdot {\mathop{ sgn}} ({r_{CG}}) + {r_{CG}} \cdot {n_{GEMM}} \cdot {\beta _G}\\
 &\ + {\alpha _G} \cdot {\mathop{ sgn}} ({r_{GG}}) + {r_{GG}} \cdot {n_{GEMM}} \cdot {\beta _G}.
\end{array}
\end{equation}

As the problem has only one unknown variable, and the highest power is 1. The optimal $r_{CG}$ must lie on the edge points of each cases, including points that $t_{L} = t_{C2G}$, $t_{G} = t_{C2G}$, $t_{L}=t_{G}$, $\tau_{G}^{n_{l}}=\tau_{C}^{n_{l}}$, $r_{CG}=0$ and $r_{CG}=1-r_{GG}$. So the time complexity is O(1). Denote an array of these points as $X$ and a corresponding array of $t_{fin}$ as $Y$.
Then, we can obtain the optimal $r_{CG}^{*}$ and corresponding $t_{fin}^{*}$ as follows:
\begin{equation}\label{rgg}
    \begin{array}{l}
t_{fin}^{*} = \min Y,\\
r_{CG}^* = X[\arg \min Y].
\end{array}
\end{equation}

\subsubsection{GPU Memory Assignment}\label{subsub:gpumem}
Slicing weight tensors of every MLP or MoE layer by $r_{GG}$, rather than transferring an entire layer's weight tensors to the GPU, allows us to effectively hide the GPU computation overhead. Importantly, variations in $r_{GG}$ lead to different inference speeds, and higher values do not necessarily yield better performance. Consequently, we introduce a simple yet efficient algorithm to determine the value of $r_{GG}$ for each layer.

Optimizing $r_{GG}$ together with $r_{CG}$ to obtain the optimal inference speed with limit GPU memory is a feasible way, but it will increase many variables and restrictions, making the optimization problem complex. Therefore, we choose to define a set $v_i=\{i/{n_{G}},1\leq i\leq n_{G}\}$ of $r_{GG}$ and calculate its time cost by Eq.~\ref{rgg}. Then, we define the importance $s_{i}^{j,t}$ of $i_{th}$ value in the set for the $j_{th}$ layer at the $t_{th}$ iteration as
  \begin{equation}
    s_{i}^{j,t} =\frac{{ {t_{fin}}^*({r_{GG}} = {v_{pre}^{j}})-{t_{fin}}^*({r_{GG}} = {v_i})}}{{({v_i} - {v_{pre}^{j}}) \cdot {n_m}}},
  \end{equation}
where $v_{pre}^{j}$ represents $r_{GG}$ in the $j_{th}$ layer at the previous iteration, and $v_{i}\cdot n_{m}$ represents the GPU memory cost when $r_{GG}=v_{i}$. We will calculate the importance of each value and each layer at each iteration. By adhering to a greedy strategy, we choose the most important $s_{i}^{j,t}$ and update its $v_{pre}^{j}$ and continue iterating until the GPU memory is exhausted.

\subsection{Token Assignment for the Prompt Phase}\label{nzero}
The optimal slicing rates during the prompt phase should clearly differ from those in the generation phase. As the generation phase always takes longer time than prompt phase, we fix slicing rates to values optimized in the generation phase. However, this will result in poor performance in the prompt phase. To improve the efficiency in the prompt phase, we introduce a token assignment schedule within Profiler \& Solver to adjust the number of tokens executed on the CPU to reduce CPU computation and improve the overlap between different resources. We initially create CG$'$, which shares CPU memory addresses with CC but operates on the GPU. Next, we allocate $n_{g}$ tokens, originally processed by CC, to run with CG$'$, thereby reducing CPU computation.
By modifying $n_{g}$, we can reduce the CPU computation time, thereby increasing the overlap among different resources.

Tokens run with CG$'$ will be processed on the GPU, requiring the transfer of CG$'$ tensors from CPU to GPU. Here, we need to change the unknown variable from $r_{CG}$ to $n_{g}$ ($0\leq n_{g} \leq T$ where $T$ denotes the number of tokens for the input) who will affect the value of $t_{L}$, $t_{C2G}$, $t_{G}$ and $t_{C}$ by  
\begin{equation}
\renewcommand{\arraycolsep}{0pt}  
\begin{array}{ll}
t_L^{'} &= [2\cdot {\mathop{ sgn}} ({r_{CG}}) + 2\cdot {\mathop{ sgn}} ({r_{\ccname{}}}) + {\mathop{ sgn}} ({r_{GG}})]\cdot t_{Lanuch},\\
t_{C2G}^{'}&=  {\alpha _{C2G}} \cdot[{\mathop{ sgn}} ({r_{CG}}) + {\mathop{ sgn}} ({r_{\ccname{}}})]+ {n_W} \cdot {\beta _{C2G}},\\
t_G^{'} &= {\alpha _G} \cdot [{\mathop{ sgn}} ({r_{CG}}) + {\mathop{ sgn}} ({r_{GG}}) + {\mathop{ sgn}} ({r_{\ccname{}}})]\\
&\ + [T \cdot ({r_{CG}} + {r_{GG}}) + {n_{g}} \cdot {r_{\ccname{}}}]MH{\beta _G},\\
t_C^{'} &= {\alpha _C} \cdot {\mathop{ sgn}} ({r_{\ccname{}}}\cdot (T - {n_{g}}))  + (T - {n_{g}}) \cdot {r_{\ccname{}}}MH{\beta _C}.
\end{array}
\end{equation}
We can use the same method to compare the time cost at edge points to determine the optimal $n_{g}^{*}$; hence, the introduction will not be reiterated. The pipelining in the prompt phase is shown in Fig.~\ref{fig:engine}. We need to optimize $n_g$ in each prompt phase.


\section{EVALUATION}\label{sec:evaluation}
\subsection{Experimental Settings}

\begin{table}[t]
	\centering
    \addtolength{\tabcolsep}{-1.6pt}
	\caption{$\alpha$ and $\beta$ of CPU GEMM, GPU GEMM and PCIe communication for Eq.\ref{pm}. $r^2$ is also given to assess the accuracy of the performance model. Notably, lanuch time is regarded as a constant so variance is used to replace $r^2$. The time unit is seconds.  }
	\label{tab:pm}
\begin{tabular}{|ll|cc|cc|c|c|}
\hline
\multicolumn{2}{|l|}{\multirow{2}{*}{}}                 & \multicolumn{2}{l|}{GPU GEMM}                            & \multicolumn{2}{l|}{CPU GEMM}                            & \multicolumn{1}{l|}{\multirow{2}{*}{PCIe}} & \multicolumn{1}{l|}{\multirow{2}{*}{Lanuch}} \\ \cline{3-6}
\multicolumn{2}{|l|}{}                                  & \multicolumn{1}{l|}{FP16}    & \multicolumn{1}{l|}{INT4} & \multicolumn{1}{l|}{FP16}    & \multicolumn{1}{l|}{INT4} & \multicolumn{1}{l|}{}                      & \multicolumn{1}{l|}{}                        \\ \hline
\multicolumn{1}{|l|}{$\alpha$}     & \multirow{3}{*}{A} & \multicolumn{1}{c|}{1.0E-7}  & 4.7E-6                    & \multicolumn{1}{c|}{7.4E-7}  & 1.1E-5                    & 3.0E-6                                     & 4.4E-5                                       \\ \cline{1-1} \cline{3-8} 
\multicolumn{1}{|l|}{$\beta$}      &                    & \multicolumn{1}{c|}{3.2E-12} & 8.1E-13                   & \multicolumn{1}{c|}{1.6E-11} & 5.4E-12                   & 2.6E-11                                    & -                                            \\ \cline{1-1} \cline{3-8} 
\multicolumn{1}{|l|}{$r^2/\sigma$} &                    & \multicolumn{1}{c|}{0.997}   & 0.999                     & \multicolumn{1}{c|}{0.988}   & 0.998                     & 0.985                                      & 3.4E-6                                       \\ \hline
\multicolumn{1}{|l|}{$\alpha$}     & \multirow{3}{*}{B} & \multicolumn{1}{c|}{1.9E-7}  & 4.6E-6                    & \multicolumn{1}{c|}{3.4E-6}  & 1.3E-5                    & 5.8E-6                                     & 5.7E-5                                       \\ \cline{1-1} \cline{3-8} 
\multicolumn{1}{|l|}{$\beta$}      &                    & \multicolumn{1}{c|}{2.6E-12} & 6.5E-13                   & \multicolumn{1}{c|}{1.5E-11} & 6.5E-12                   & 2.5E-11                                    & -                                            \\ \cline{1-1} \cline{3-8} 
\multicolumn{1}{|l|}{$r^2/\sigma$} &                    & \multicolumn{1}{c|}{0.997}   & 0.996                     & \multicolumn{1}{c|}{0.995}   & 0.998                     & 0.994                                      & 5.9E-6                                       \\ \hline
\multicolumn{1}{|l|}{$\alpha$}     & \multirow{3}{*}{C} & \multicolumn{1}{c|}{1.4E-7}  & 6.4E-6                    & \multicolumn{1}{c|}{1.8E-6}  & 5.6E-7                    & 3.7E-6                                     & 5.2E-5                                       \\ \cline{1-1} \cline{3-8} 
\multicolumn{1}{|l|}{$\beta$}      &                    & \multicolumn{1}{c|}{3.6E-12} & 9.2E-13                   & \multicolumn{1}{c|}{2.5E-11} & 8.4E-12                   & 4.1E-11                                    & -                                            \\ \cline{1-1} \cline{3-8} 
\multicolumn{1}{|l|}{$r^2/\sigma$} &                    & \multicolumn{1}{c|}{0.988}   & 0.989                     & \multicolumn{1}{c|}{0.993}   & 0.992                     & 0.999                                      & 6.0E-6                                       \\ \hline
\end{tabular}
\end{table}

\textbf{Testbeds.}: Experiments are carried out on three testbeds: \textit{Testbed-A}, Nvidia RTX A6000 GPU, Intel(R) Xeon(R) Platinum 8358 CPU and PCIe-4.0x16. \textit{Testbed-B}, Nvidia RTX 3090 GPU, AMD EPYC 7742 and PCIe-4.0x16. \textit{Testbed-C}, Nvidia RTX 2080Ti GPU, Intel(R) Xeon(R) Gold 6230 CPU and PCIe-3.0x16. The software environments are Ubuntu-22.04, CUDA-12.1 and PyTorch-2.1.2.

\textbf{Models.} We use both dense and sparse models including LLaMA\cite{DBLP:journals/corr/abs-2302-13971}, Qwen\cite{DBLP:journals/corr/abs-2407-10671}, Mixtral\cite{jiang2024mixtral} and PhiMoE\cite{DBLP:journals/corr/abs-2404-14219}. All models in our experiments use quantized parameters of FP16 or INT4.

\textbf{Baseline Systems.} We use Fiddler and llama.cpp as our baselines. Fiddler is particularly optimized for MoE models and llama.cpp is a well-known inference system that utilizes both CPU and GPU resources to accelerate inference.

\textbf{Workloads.} The workloads for our experiments are derived from the ChatGPT prompts\footnote{https://huggingface.co/datasets/MohamedRashad/ChatGPT-prompts}. 

\subsection{Performance Models}\label{subsec:perfmodel-evaluation}
We measure the launch time of GPU kernels and the elapsed time with a range of sizes for CPU and GPU GEMM operations and the communication between GPU and CPU to fit the performance models in Eq.~\ref{pm}. As we denote the workload of a GEMM as $n_{GEMM}=T \cdot M \cdot H$, we need to measure two sets of $\alpha_{C}, \beta_{C}, \alpha_{G}, \beta_{G}$ for each testbed. We also provide $r^2$ and $\sigma$ to check the accuracy of the performance model. The results are shown in Table~\ref{tab:pm}, which indicates that our linear models with intercept terms (i.e., startup time) can well fit the measured performance. 

\begin{figure}[!t]
        \begin{subfigure}[t]{0.48\linewidth}
		\centering
		\includegraphics[width=1.0\textwidth]{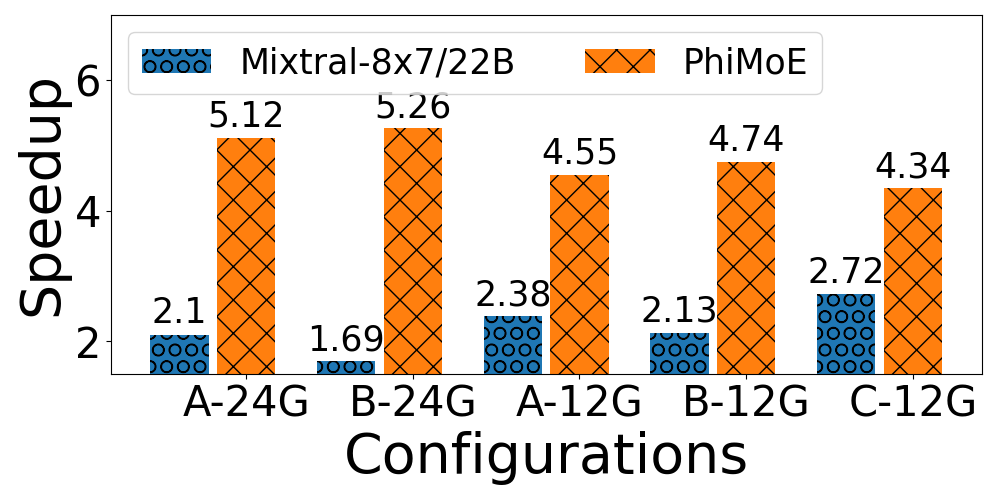}
		\caption{prompt phase.}
		\label{fig:prefill-fiddler}
	\end{subfigure}
	\begin{subfigure}[t]{0.48\linewidth}
		\centering
		\includegraphics[width=1.0\textwidth]{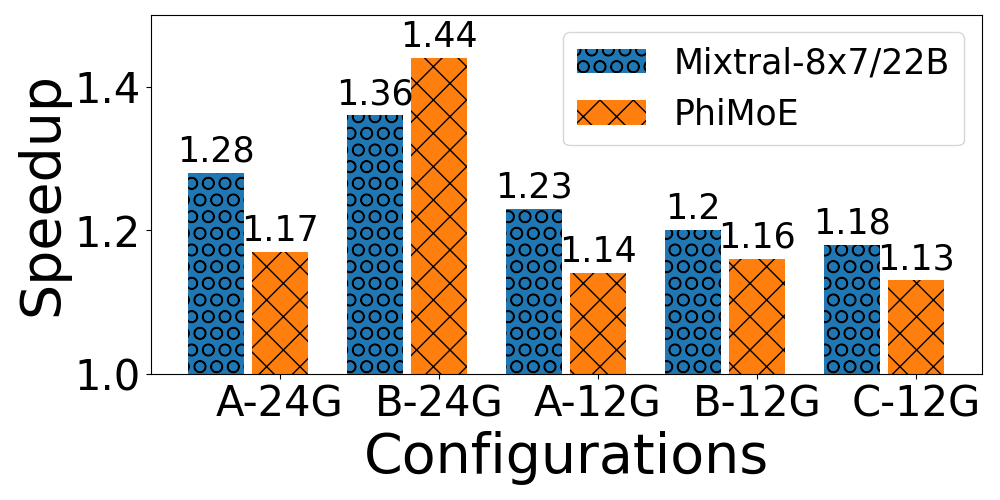}
		\caption{generation phase.}
		\label{fig:decode-fiddler}
	\end{subfigure}

	\caption{The speedups of \modelname{} over Fiddler with different models, testbeds and different GPU memory constraints.}
	\label{fig:fiddler-exp}
\end{figure}

\begin{figure}[!t]
        \begin{subfigure}[t]{\linewidth}
		\centering
		\includegraphics[width=1.0\textwidth]{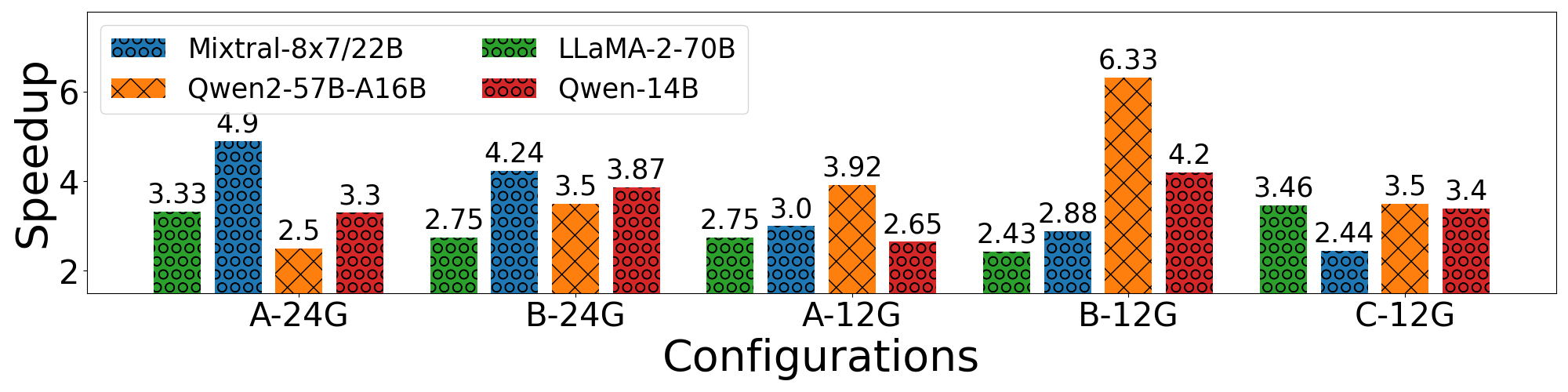}
		\caption{prompt phase.}
		\label{fig:prefill-llamacpp}
	\end{subfigure}
    
	\begin{subfigure}[t]{1.05\linewidth}
		\centering
		\includegraphics[width=1.0\textwidth]{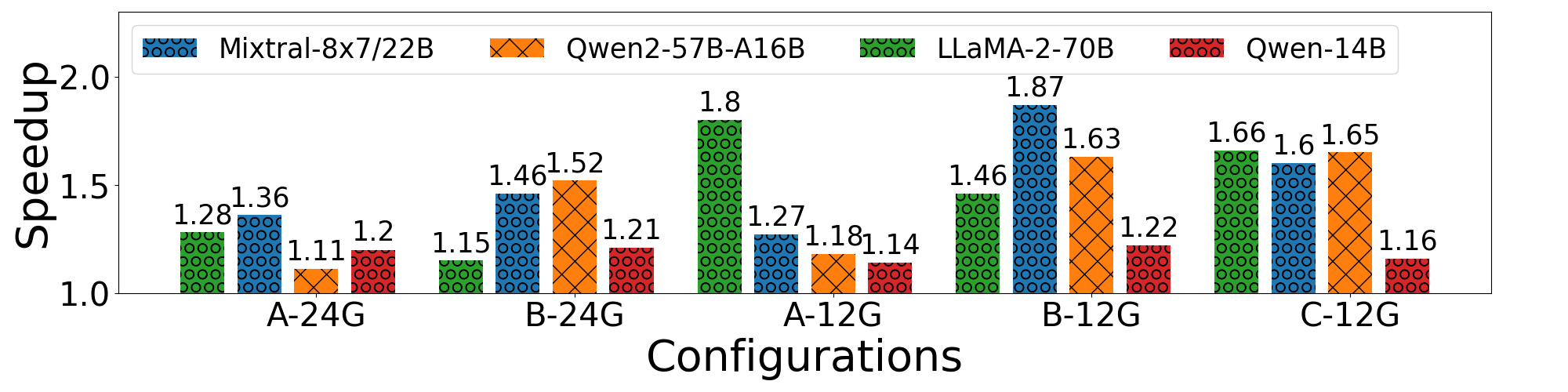}
		\caption{generation phase.}
		\label{fig:decode-llamacpp}
	\end{subfigure}

	\caption{The speedups of \modelname{} over llama.cpp with different models, testbeds and different GPU memory constraints.}
	\label{fig:llamacpp-exp}
\end{figure}

\subsection{Performance on the Prompt Phase}
We evaluate the speedup of our \modelname{} over Fiddler and llama.cpp with 4bit quantized LLaMA-2-70B, Qwen2-57B-A14B, Mixtral-7B and Mixtral-22B and FP16 PhiMoE, LLaMA-2-13B and Qwen2-14B. The input length is set to 1024 by default. Notably, Mixtral-22B is used only when the GPU memory is 24GB. The results are shown in Fig.~\ref{fig:prefill-fiddler} and Fig.~\ref{fig:prefill-llamacpp}, which show that \modelname{} achieves speedups ranging from $1.69\times$ to $6.33\times$ over Fiddler and llama.cpp in the prompt phase. 

\textbf{Comparison with PowerInfer.} PowerInfer~\cite{song2023powerinfer} represents another optimized inference system that streamlines the model by substituting the original activation functions with ReLU. This modification could influence the outcomes produced during the generation stage; therefore, we restrict our comparison to the prompt phase. We conduct experiments using the 4-bit quantized LLaMA-2-70B and FP16 LLaMA-2-13b on testbeds A and B, both have a 12GB memory constraint. The experimental results show that \modelname{} accelerates the prompt performance by speedups ranging from $6.9\times$ to $19.2\times$.

\subsection{Performance on the Generation Phase}
Performance comparison in the generation phase is shown in Fig.~\ref{fig:decode-fiddler} and Fig.~\ref{fig:decode-llamacpp}. The results show that \modelname{} achieves speedups ranging from $1.11\times$ to $1.87\times$ over Fiddler and llama.cpp. The time cost of the communication between CPU to GPU is much larger than CPU GEMM which is also much larger than GPU GEMM in the same workload on our testbeds in the generation phase. Consequently, the improvement is not as pronounced as that observed during the prompt phase.

Putting prompt and generation phases together, \modelname{} achieves end-to-end inference performance improvements by $1.25\times$ to $2.04\times$ over llama.cpp and Fiddler.

\subsection{Ablation Study}
\subsubsection{Impacts of the CPU GEMM Speed}
To understand the impacts of the CPU GEMM speed, we configure the number of threads in the range of [16, 8, 4, 2] in executing CPU GEMM in the generation phase. The results shows that our \modelname{} achieves average speedups of $1.23\times$, $1.29\times$, $1.43\times$, and $2.4\times$ over Fiddler while $1.45\times$, $1.48\times$, $1.63\times$, and $2.09\times$ over llama.cpp. It shows that \modelname{} is more effective on lower-performance CPUs.
\subsubsection{Importance of Slicing Rates}\label{subsubsec:abla_rate}
To assess the significance of $r_{CG}$, we measure the performance in the generation phase using the optimal slicing rates $r_{CG}$ against those with constant slicing rates of 0, 0.25, 0.5, and 0.75 on 4-bit quantized Mixtral-7B and LLaMA-2-70B for Testbed A, considering the memory constraints of 12GB and 24GB, respectively. Note that we set $r_{GG}=0$ to disable the GPU memory assignment schedule and exclude its impact. Furthermore, we conduct experiments using a varying number of CPU threads from 16 down to 2 to check its ability on different CPU GEMM speeds. The experimental results show that our optimal slicing rates result in average speedups of $1.7\times$, $1.4\times$, $1.6\times$, and $2.1\times$ compared to the fixed slicing rates, verifying the significance of optimal slicing rates. Our optimal slicing rate varies with different CPU threads. For instance, with 16, 8, 4, and 2 CPU threads, the optimal slicing rates are 0.18, 0.25, 0.42, and 0.71 on LLaMA-2-70B, respectively.
\subsubsection{Effect of the GPU Memory Assignment}
To evaluate how the GPU memory assignment schedule affects \modelname{}, we conduct a performance comparison of our \modelname{} w/ and w/o the GPU memory assignment. Specifically, when the GPU memory assignment is disabled, we set $r_{GG}$ to 0 or 1 and adjust the number of layers with $r_{GG}=1$ to maximize GPU memory utilization. Testing was conducted on a 4-bit quantized Mixtral-7B, adhering to memory constraints of 12, 16, and 20GB, as well as a 4-bit quantized LLaMA-2-70B with memory constraints of 24, 28, and 32GB for Testbed A. Our findings reveal that GPU memory assignment can result in speed improvements of $1.07 \times$, $1.10 \times$, and $1.15 \times$ for Mixtral-7B, and $1.08 \times$, $1.09 \times$, and $1.12 \times$ for LLaMA-2-70B, substantiating the effectiveness of GPU memory assignment.
\subsubsection{Necessity of the Token Assignment in the Prompt Phase}
To verify the importance of our token assignment schedule during the prompt phase, we perform a performance comparison of our \modelname{} both with and without token assignment. We conduct experiments using sequence lengths of 64, 256, and 1024 on a 4-bit quantized Mixtral-7B, maintaining a memory limit of 12GB, as well as on a 4-bit quantized LLaMA-2-70B with a 24GB memory limit for Testbed A. The findings indicate that token assignment leads to speed boosts of $1.9 \times$, $5.6 \times$, and $19.1 \times$ for Mixtral-7B, and $6.9 \times$, $27.6 \times$, and $45.6 \times$ for LLaMA-2-70B, underscoring the need for token assignment.
\section{Related Works}
\textbf{LLM Inference Optimizations under Memory Constraints}: To enable inference on local platforms like desktop computers with limited computational power and memory capacity, recent researches have tried to sparse weights by prunig\cite{han2015deep,han2015learning,ma2023llm}, offloading weights between the CPU and the GPU \cite{kamahori2024fiddler,llamacpp,song2023powerinfer，he2024expertflowoptimizedexpertactivation}, employing flash memory\cite{alizadeh2023llm} or using the CPU for the computation\cite{llamacpp,kamahori2024fiddler}. However, these works still struggle with the utilization of different resources. In contrast, \modelname{} seeks to utilize CPU, GPU computation, and PCIe communication simultaneously to improve performance.

\textbf{LLM Attention Optimizations}: Computing LLM attention requires storing a key value cache, and its workload grows quadratically with the input sequence length. To enhance attention performance for longer sequences, Sglang\cite{zheng2023sglang} and Hydragen\cite{juravsky2024hydragen} maximize computational sharing between sequences. vLLM \cite{kwon2023efficient} proposes paged memory methods to efficiently manage the key value cache. Although this is beyond our current focus, incorporating these attention optimizations into \modelname{} could potentially accelerate LLM inference.

\textbf{LLM MLP or MoE Optimizations}: With moderate sequence lengths, LLM MLP or MoE tend to consume more memory and computational resources than attention, particularly in MoE models. To mitigate MoE's resource impact, Fiddler\cite{kamahori2024fiddler} suggests storing experts on the CPU to preserve GPU memory. AdapMoE\cite{zhong2024adapmoe} introduces a system to manage hot and cold experts. ExpertFlow\cite{he2024expertflowoptimizedexpertactivation} presents a predictive routing path-based offloading strategy to maintain expert caching.. Some activation sparsity techniques\cite{song2023powerinfer,wang2024q,DBLP:conf/icml/LiuWDZY0S0TRC23} attempt to predict activation sparsity to decrease computation and storage for MLPs or MoEs. However, none of these approaches integrates CPU, GPU, and PCIe communication resources effectively to optimize utilization. Furthermore, the sparsity of activation will certainly affect LLM outputs.

\section{Conclusion}
In this work, we proposed \modelname{}, which is an inference system designed for efficient LLM inference on computer systems with a single moderate GPU. \modelname{} coordinates CPU, GPU, and PCIe communication tasks to utilize available computing resources efficiently, thus improving the inference speed. The experimental results show that \modelname{} outperforms existing solutions, Fiddler and llama.cpp, providing a $1.11\times$ to $1.80\times$ speed increase in the generation phase and $1.69\times$ to $6.33\times$ in the prompt phase, with Mixtral, LLaMA-2, Qwen, and PhiMoE models in three testbeds.
\bibliography{ref}
\bibliographystyle{IEEEtran}
\end{document}